\documentclass[letterpaper, twocolumn, pra]{revtex4}
\usepackage{graphicx, amsmath, amssymb, bm, titlesec, mathtools}

\begin{document}

\title{Experimental demonstration of structural robustness of spatially partially coherent fields in turbulence}

\author{Abhinandan Bhattacharjee and Anand K. Jha}

\email{akjha9@gmail.com}

\affiliation{Department of Physics, Indian Institute of
Technology Kanpur, Kanpur UP 208016, India}

\date{\today}
\begin{abstract}

Structured fields that are spatially completely coherent have been extensively studied in the context of long-distance optical communication as the structure in the intensity profile of such fields is used for encoding information. This method of doing optical communication works very well in the absence of turbulence. However, in the presence of turbulence, the intensity structures of such fields start to degrade because of the complete spatial coherence of the field, and this structural degradation increases with the increase in the turbulence strength. On the other hand, several theoretical studies have now shown that the structured fields that are spatially only partially coherent are less affected by turbulence. However, no such experimental demonstration has been reported until now. In this letter, we experimentally demonstrate the structural robustness of partially coherent fields in the presence of turbulence, and we show that for a given turbulence strength the structural robustness of a partially coherent field increases as the spatial coherence length of the field is decreased. 

\end{abstract}

\maketitle


In the past few decades, structured fields that are spatially completely coherent, such as Laguerre Gaussian (LG) and Hermite Gaussian (HG) modes produced by stable laser resonators \cite{forbes2019structured} or spatial light modulators (SLMs) \cite{matsumoto2008josaa}, have gained importance due to their implications for optical communication \cite{andrews2005laser,majumdar2010springer,lavery2017sciadv,ren2016scirpt,trichili2016optlet,aksenov2013jop,yuan2010optnlt,ndagano2017optlet,cox2018prapplied,pang2018optlet}.  
The structure in the intensity profile of such fields is used for encoding information in the long-distance fiber \cite{zhu2016optexp} and free space \cite{krenn2014njp,krenn2016pnas,erhard2018LSA} optical communication.
However, the problem in using such structured fields in the presence of a turbulent medium is that the medium introduces random phase fluctuations at different
spatial locations in the field, and due to the perfect spatial coherence of the field these random phase fluctuations result in the degradation of the intensity structures of such fields. 
As a consequence, the retrieval of information encoded in the intensity structures becomes difficult. 
Due to this reason, the structures in a spatially perfectly coherent field become unsuitable for encoding and transferring information in turbulent environments.

On the other hand, it is now known that a spatially partially coherent field is less affected by turbulence  \cite{roychowdhury2004oc,avramov2016oc,ricklin2002josaa,salem2003oc,wang2017ao}. Furthermore, theoretical studies have now shown that in the presence of turbulent environments
the structures in the intensity profiles and in the cross-spectral density functions of a spatially partially coherent field degrade less in comparison to the intensity structures of a spatially perfectly coherent field \cite{mei2013optexp,yu2018optexp,zhu2017optexp,liu2016optlet,lin2020optexp}. 
This implies that the structural robustness of the intensity profiles and the cross-spectral density functions of a spatially partially coherent field could be utilized towards optical communication  even in the presence of a turbulent environment. 
Although in the past few years,
there has been a growing interest in engineering various structured
fields that are spatially partially coherent \cite{liang2018optlet,chen2016apl,chen2014optexp,cai2014josaa,liang2014optlet}, to the best of our knowledge, no experimental demonstration of structural robustness of the cross-spectral density function of such fields in turbulence has been reported so far.
\begin{figure}[b!]
\includegraphics[scale=1.00]{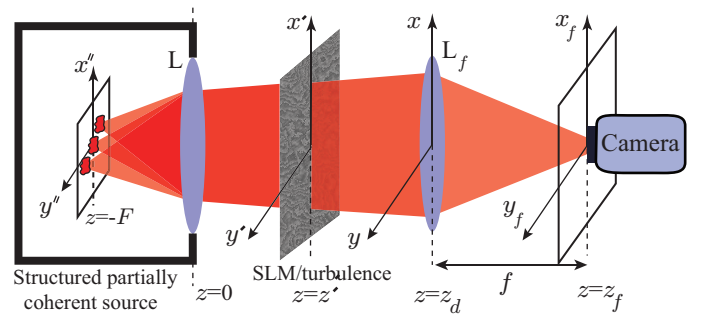}
\caption{(Colour online) Schematic of the experimental setup illustrating propagation of our structured partially coherent source through a turbulent medium.}\label{fig1}
\end{figure}
In this letter, we experimentally demonstrate structural robustness of partially coherent fields in turbulent environments. Simulating planar turbulence with the help of an SLM, we show that for a given turbulence strength the structural robustness of a partially coherent field increases as the spatial coherence length of the field is decreased. 


Figure~\ref{fig1} shows the schematic of our experimental setup and also illustrates how our structured partially coherent source propagates through a planar simulated turbulence and gets detected. In our experimental demonstrations, we use the scheme of Ref.~\cite{aarav2017pra} for generating spatially partially coherent fields with structures in their cross-spectral density functions. A planar, monochromatic, spatially completely incoherent primary source is kept at the back focal plane $z=-F$ of a lens located at $z=0$. The central wavelength of the source is $\lambda_0=2\pi/k_0$, where $k_0$ is the magnitude of the wavevector. The combination of the primary incoherent source along with the lens constitute our structured spatially partially coherent source. The structured partially coherent field passes through a planar simulated turbulence kept at $z=z'$ and then gets observed by the detection system consisting of a converging lens of focal length $f$ kept at $z=z_d$ and a camera kept at $z=z_f=z_d+f$. The detection system essentially measures the cross-spectral density function of the field at $z=z_d$. We represent the transverse position coordinates at $z=-F$, $z=z'$, $z=z_d$, and $z=z_f$ by $\bm{\rho}''\equiv (x'',y'')$, $\bm{\rho}'\equiv (x',y')$, $\bm{\rho}\equiv (x,y)$, and $\bm{\rho}_f \equiv (x_f,y_f)$, respectively. The intensity of the primary source at $z=-F$ is given by $I (\bm{\rho}'', z=-F)$. Therefore the cross-spectral density function $W_s(\rho_1', \rho_2'; z=z')$ of our partially coherent field at $z=z'$ can be shown to be \cite{aarav2017pra}  
\begin{figure}[t!]
\includegraphics{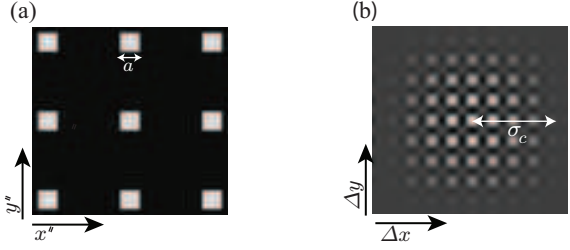}
\caption{(Colour online) (a) Simulated intensity of the primary source. (b) Simulated cross-spectral density $W_s(\Delta\rho;z=z_d)$ of the source at $z=z_d$}\label{fig2}
\end{figure}
\begin{align}
W_s(\bm{\rho}'_1,\bm{\rho}'_2;z=z') \rightarrow W_s(\Delta\rho';z=z') \notag \\
=\tfrac{A}{F^2}\iint I(\bm{\rho}'';z=-F) e^{-i\tfrac{k_0}{F}\bm{\rho}''\cdot\Delta\bm{\rho'}}d\bm{\rho}''\label{cross-spectral-1}
\end{align}
where $\Delta{\rho}'=|\bm{\rho'_2}-\bm{\rho'_1}|$. We note that the cross-spectral density function $W_s(\bm{\rho}'_1, \bm{\rho}'_2; z=z')$ of our source depends on the transverse coordinates only through their difference $\Delta\rho'$. Therefore, we write it as $W_s(\Delta\rho'; z=z')$. Such sources are referred to as statistical homogeneous source \cite{mandel1995coherence} or even spatially stationary source \cite{aarav2017pra}. 
The cross-spectral density $W(\bm{\rho'_1},\bm{\rho'_2};z=z')$ at $z=z'$ right after the turbulence plane is given by $W(\bm{\rho'_1},\bm{\rho'_2};z=z')=W_s(\Delta\rho';z=z')W_{t}(\bm{\rho'_1},\bm{\rho'_2})$, where $W_{t}(\bm{\rho'_1},\bm{\rho'_2})$ is the cross-spectral density induced due to the turbulence. According to the Kolmogorov model, 
\begin{align}
W_{t}(\bm{\rho'_1},\bm{\rho'_2})=e^{-3.44(\frac{\Delta{\rho'}}{r_0})^{\tfrac{5}{3}}}.\label{turb}  
\end{align}
The quantity ${r_{0}}$ is called Fried's coherence diameter \cite{fried1966josaa,boyd2011optexp}, and it quantifies the strength of turbulence. The value of $r_0$ ranges from $0$ to $\infty$, with limit $r_0 \rightarrow 0$ implying infinite turbulence strength and limit $r_0 \rightarrow \infty$ implying no turbulence. In order to show the structural robustness of our partially coherent field in turbulence, we obtain the cross-spectral density function of the field after it has propagated up to $z=z_d$. Using Eqs.~(\ref{cross-spectral-1}) and ~(\ref{turb}) and the Wolf propagation equation $($section 4.4.3 of Ref.~\cite{mandel1995coherence}$)$,  we find the cross spectral density function $W(\bm{\rho_1},\bm{\rho_2};z=z_d) \rightarrow W(\Delta\rho;z=z_d)$ of the field at $z=z_d$ to be
\begin{align}
W(\Delta\rho;z=z_d)=e^{-3.44(\tfrac{\Delta\rho}{r_0})^{\frac{5}{3}}}W_s(\Delta\rho;z=z_d).\label{eq5}  
\end{align}
where  
\begin{align}
W_s(\Delta\rho;z=z_d)=\tfrac{A}{F^2}\iint{I(\bm{\rho}'';z=-F)e^{-i\tfrac{k_0}{F}\bm{\rho}''\cdot\Delta\bm{\rho}}d\bm{\rho}''}\label{cross-spectral}
\end{align}
is the cross-spectral density function of the field at $z=z_d$ in the absence of turbulence, and $\Delta\rho=|\bm{\rho}_2-\bm{\rho}_1|$. We note that the cross-spectral density functions $W(\Delta\rho;z=z_d)$ and $W_s(\Delta\rho;z=z_d)$ depend on the transverse position coordinates only through their difference $\Delta\rho$ and thus that the field at $z=z_d$ remains spatially stationary with or without the turbulence. Furthermore, we note that $W_s (\Delta\rho; z=z_d)$ remains propagation invariant \cite{aarav2017pra} and therefore it has the same functional form as that of the cross-spectral density function $W_s (\Delta\rho; z=z')$ at $z=z'$, as given in Eq.~(\ref{cross-spectral-1}). 
We note that since $W(\Delta\rho; z=z_d)$ is spatially stationary, it can be expressed in terms of the intensity $I (\rho_f; z=z_f)$ at $z=z_f$. In order to show this we first write the cross-spectral density $W_{l}(\bm{\rho_1},\bm{\rho_2},z_d)$ at $z=z_d$ right after the lens $L_f$ as $W_{l}(\bm{\rho_1},\bm{\rho_2};z=z_d)=W(\Delta\rho;z=z_d)T^{*}(\bm{\rho_1})T(\bm{\rho_2})$, where $T(\bm{\rho})=e^{i\tfrac{k_0}{2f}\rho^2}$ is the transmission function of lens $L_f$ \cite{goodman2005introduction}. Next, using the Wolf propagation equation \cite{mandel1995coherence}, we propagate the field from $z=z_d$ to $z=z_f$ and find the intensity $I(\bm{\rho_f};z=z_f)$ at $z=z_f$ plane to be
\begin{figure}
\includegraphics[scale=1.00]{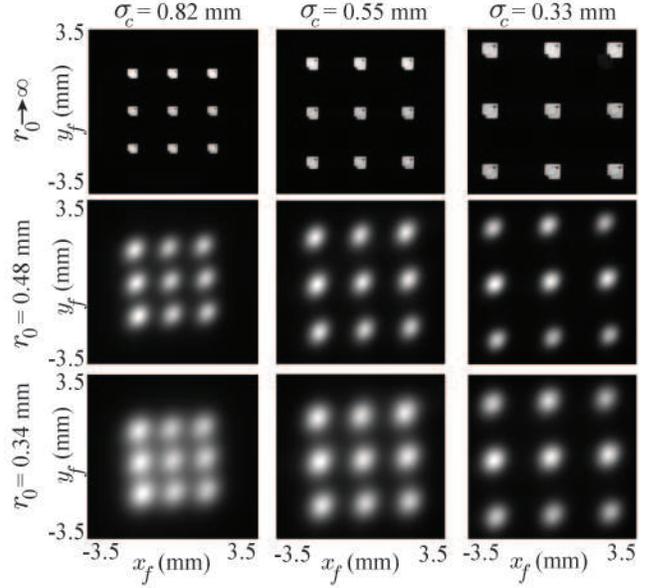}
\caption{(Colour online) Experimentally measured $I({\bm{\rho_f}};z=z_f)$ with different transverse coherence lengths at various turbulence strengths.}\label{fig3}
\end{figure}
\begin{align}
I({\bm{\rho_f}};z=z_f)=&W({\bm{\rho_f}},{\bm{\rho_f}};z=z_f) \notag \\
&=\iint W(\Delta\rho;z=z_d)e^{i\tfrac{k_0}{f}{\bm{\rho_f}}\cdot\Delta\bm{\rho}}d{\Delta\bm{\rho}}\label{fourier}
\end{align}
We rewrite the above equation as 
\begin{align}
W(\Delta{{\rho}};z=z_d)=\iint I({\bm{\rho_f}};z=z_f)e^{-i\tfrac{k_0}{f}{\bm{\rho_f}}\cdot\Delta\bm{\rho}}d{\bm{\rho_f}}\label{fourier1}
\end{align}
Thus we see that by measuring the intensity $I(\rho_f; z=z_f)$ at the focal plane $z=z_f$, one obtains the cross-spectral density function $W(\Delta\rho, z=z_d)$ at $z=z_d$. 

\begin{figure*}[t!]
\centering
\includegraphics[scale=1.00]{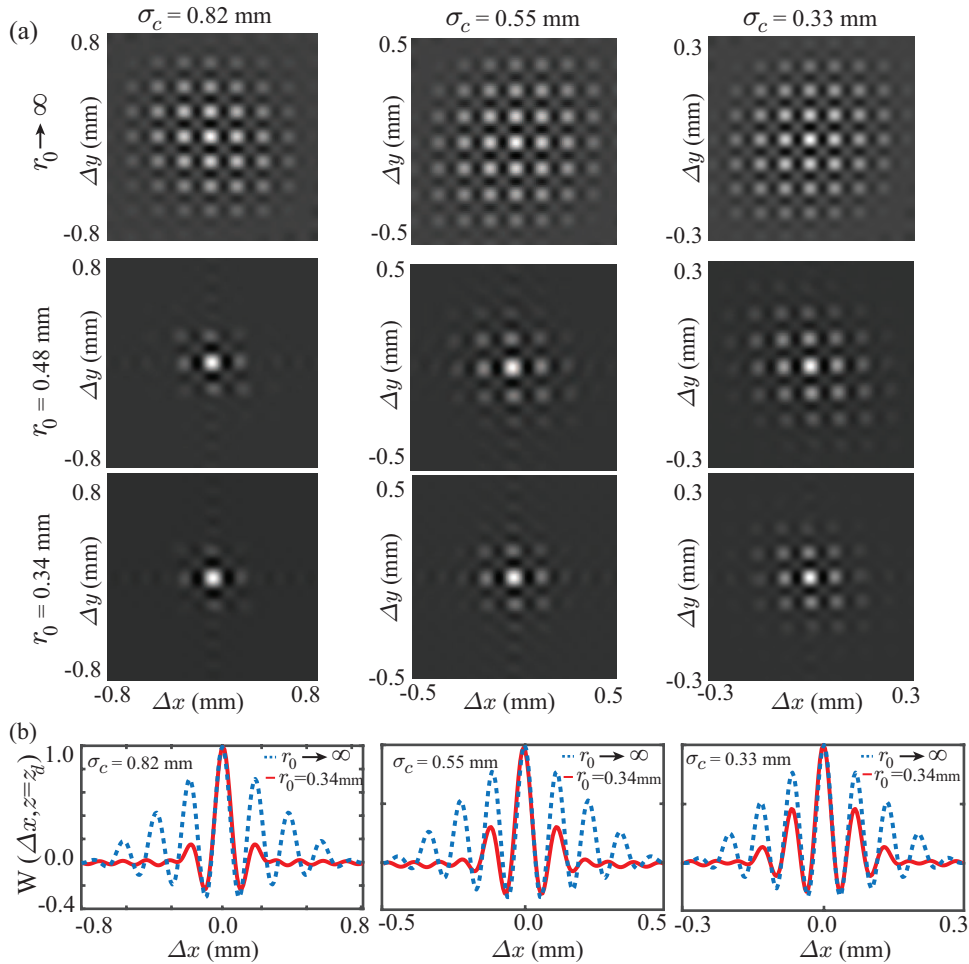}
\caption{(Colour online) (a) Reconstructed cross-spectral density function $W(\Delta\rho;z=z_d)$ for different transverse coherence lengths at various turbulence strengths. (b) The plots of one-dimensional cuts along the $x-$direction of $W(\Delta\rho;z=z_d)$ at $r_0 \rightarrow \infty$ and $r_0=0.34$ mm for different $\sigma_c$.}\label{fig5}
\end{figure*}

We next present our experimental demonstration of structural robustness of spatially partially coherent fields in the presence of turbulent media. 
Figure~\ref{fig1} shows the schematic of the experimental setup, where the structured partially coherent source is kept at $z=0$. We use a spatial light modulator (SLM) for simulating planar turbulence at $z=z'$ \cite{rodenburg2012optlet} and an electron multiplied charged coupled device (EMCCD) camera for measuring the intensity at $z=z_f$ plane. From Eq.~(\ref{cross-spectral}), we have that the cross-spectral density function $W_s(\Delta\rho;z=z_d)$ at $z=z_d$ is the Fourier transform of the intensity $I(\bm{\rho}'';z=-F)$ of the primary incoherent source. Therefore, in order to generate spatially partially coherent field with structured cross-spectral density function, we use a light emitting diode (LED) array as our primary source. The array consists of $9$ LEDs arranged in a $3\times3$ grid. The size of the individual LED is $a = 0.58$ mm. Figures~\ref{fig2}(a) shows the simulated intensity $I (\bm{\rho}''; z=-F)$ of our primary incoherent source at $z=-F$ while Fig. 2(b) shows the corresponding cross spectral density function $W_s(\Delta\rho; z=z_d)$ at $z=z_d$. We note that the oscillatory features of the cross-spectral density function in Fig.~\ref{fig2}(b) decays over a length scale $\sigma_c$ in the transverse direction. Using Eq.~(\ref{cross-spectral}), it can be shown that $\sigma_c$ is decided by the transverse size $a$ of the individual LEDs at $z=-F$ and that it can be written as $\sigma_c= \lambda_0 F/a$ (see Ref.~\cite{mandel1995coherence}, section 4.4.4). We take $\sigma_c$ as the spatial coherence length of the field. This definition of the spatial coherence length is consistent with the definition of temporal coherence length for a multi-mode continous wave (CW) laser with structured temporal cross-spectral density function \cite{mandel1962measures}. By using lenses of focal lengths $F = 30$ cm, $50$ cm, and $75$ cm in the source configuration, we generate structured spatially partially coherent fields with $\sigma_c= 0.33$ mm, $0.55$ mm, and $0.82$ mm, respectively. 
In order to simulate turbulence using an SLM kept at $z=z'$, we display around $200$ random phase patterns on the SLM with Kolmogorov statistics in a sequential manner at a frame rate of $30$ Hz. We set an exposure time of $7$ seconds such that the  EMCCD camera records the entire ensemble of fields corresponding to the 200 phase patterns. In this way, we generate Kolmogorov turbulence. We perform experiments at three different turbulence strengths $r_0 \rightarrow \infty$, $r_0=0.48$ mm, and $r_0=0.34$ mm.    




In our experiments, we use $f=30$ cm, $z'=20$ cm, $z_d=50$ cm, and $z_f=z_d+f=80$ cm. Figure~\ref{fig3} shows the experimentally measured intensity $I({\bm{\rho_f}};z=z_f)$ at $z=z_f$ for different spatial coherence lengths $\sigma_c$ at various turbulence strengths $r_0$. 
With no turbulence, that is, at $r_0 \rightarrow \infty$, the intensity $I(\bm{\rho}_f;z=z_f)$ at different $\sigma_c$ is the same as the intensity $I(\bm{\rho}'';z=-F)$ of the primary source shown in Fig.~\ref{fig2}(a), apart
from a change in scale. In the presence of turbulence, we find that as the spatial coherence length $\sigma_c$ of the field decreases from $0.82$ mm to $0.33$ mm, the degradation in the structural features of the intensity $I(\bm{\rho}_f;z=z_f)$ becomes lesser. The small tilt in the measured intensity of Fig.~\ref{fig3} is attributed to the imperfections in the alignment of the experimental setup. 
 
Next, using Eq.~(\ref{fourier1}), we reconstruct the cross-spectral density function $W(\Delta\rho;z=z_d)$ at $z=z_d$ from the above measured intensity $I({\bm{\rho_f}};z=z_f)$. Figure~\ref{fig5}(a) shows the reconstructed cross-spectral density function $W(\Delta\rho;z=z_d)$ at $z=z_d$ for different $\sigma_c$ at various $r_0$. 
We see that in the absence of turbulence, that is at $r_0 \rightarrow \infty$, the two-dimensional structure profile of $W(\Delta\rho;z=z_d)$ is same for all three $\sigma_c$ values, apart from a change in scale. In the presence of turbulence, we find that the two-dimensional structures suffer degradation for all three $\sigma_c$ values. However, at a given turbulence strength, the structural degradation becomes less as the spatial coherence length is decreased. 
We note that in Fig.~\ref{fig5}(a), we have plotted $W(\Delta\rho;z=z_d)$ over different range of $\Delta\rho=(\Delta x, \Delta y)$ at different $\sigma_c$. This is so that we can better compare the structural degradation at different $\sigma_c$ values. Finally, in order to highlight the main claim of this letter, which is that the structural robustness increases as $\sigma_c$ in decreased, we plot in Fig.~\ref{fig5}(b) the one-dimensional cross-spectral density function  $W(\Delta x; z=z_d)$ by taking one-dimensional cuts of $W(\Delta\rho; z=z_d)$ plots in Fig.~\ref{fig5}(a). For each $\sigma_c$, we plot $W(\Delta x; z=z_d)$ at $r_0 \rightarrow \infty$, and $r_0=0.34$ together. These plot clearly show that the structural robustness of the cross-spectral density function of a spatially partially coherent field increases as we decrease the spatial coherence length of the field.  

In conclusion, we have experimentally demonstrated structural robustness of spatially partially coherent fields in the presence of turbulence. We have shown that at a given turbulence strength the structural robustness of a partially coherent field increases with the decrease in the spatial coherence length of the field. Our work can have important implications for long-distance optical communication through turbulent environments. We note that in our experiments, we have worked with simulated planar turbulence of strength $r_0$ ranging from $\infty$ to $0.34$ mm. On the other hand, the real atmospheric turbulence is distributed and can even cause amplitude fluctuations in addition to random  phase fluctuations. The typical values of $r_0$ for atmospheric turbulence range from $4$ mm to $30$ mm \cite{krenn2016pnas,krenn2014njp,lavery2017sciadv}. So, although there are some basic differences between the real atmospheric turbulence and the planar turbulence used in our experiments, we expect the main result of this letter to remain qualitatively valid even for the real atmospheric turbulence. We further note that the scheme presented in this letter for measuring the cross-spectral density function works only for spatially-stationary partially coherent fields. However, there are non spatially-stationary partially coherent fields \cite{ponomarenko2007ol,zhu2019ol} that have been found to have very interesting propagation properties in turbulence \cite{li2020ol}. We expect that the main result of this letter will remain valid even for non spatially-stationary fields, although in that case one would need to use  a different scheme for measuring the cross-spectral density function \cite{naraghi2017ol}.

We thank Shaurya Aarav for useful discussions and acknowledge financial support through the research grant no. EMR/2015/001931 from the Science and Engineering Research Board (SERB), Department of Science and Technology, Government of India and through the research grant no. DST/ICPS/QuST/Theme- 1/2019 from the Department of Science and Technology, Government of India.


\begin{thebibliography}{10}

\bibitem{forbes2019structured}
Andrew Forbes, Laser \& Photonics Reviews, {\bf 13}, 1900140 (2019).

\bibitem{matsumoto2008josaa}
Naoya Matsumoto, Taro Ando, Takashi Inoue, Yoshiyuki Ohtake, Norihiro Fukuchi, and Tsutomu Hara, JOSA A, {\bf 25}, 1642 (2008).

\bibitem{andrews2005laser}
Larry C Andrews, and Ronald L Phillips, {\em Laser beam propagation through random media}, ({SPIE press Bellingham, WA} 2005).

\bibitem{majumdar2010springer}
Arun K Majumdar,  and Jennifer C Ricklin, {\em Free-space laser communications: principles and advances}, {\bf {Vol.2}}, ({Springer Science \& Business Media 2010}).

\bibitem{lavery2017sciadv}
Martin PJ Lavery, Christian Peuntinger, Kevin G{\"u}nthner, Peter Banzer, Dominique Elser, Robert W Boyd, Miles J Padgett, and Christoph Marquardt, and Gerd Leuchs, Science advances, {\bf 3}, e1700552 (2017).

\bibitem{ren2016scirpt}
Yongxiong Ren, Long Li, Zhe Wang, Seyedeh Mahsa Kamali, Ehsan Arbabi, Amir Arbabi, Zhe Zhao, Guodong Xie, Yinwen Cao, Nisar Ahmed, Yan Yan, Cong Liu, Asher J. Willner, Solyman Ashrafi, Moshe Tur, Andrei Faraon, and Alan E. Willner, Scientific reports, {\bf 6}, 33306 (2016).

\bibitem{trichili2016optlet}
Abderrahmen Trichili, Amine Ben Salem, Angela Dudley, Mourad Zghal, and Andrew Forbes, Optics letters, {\bf 41}, 3086 (2016).

\bibitem{aksenov2013jop}
VP Aksenov, VV Kolosov, and CE Pogutsa, Journal of Optics, {\bf 15}, 044007 (2013)

\bibitem{yuan2010optnlt}
Yangsheng Yuan, Yangjian Cai, Jun Qu, Halil T Eyyubo{\u{g}}lu, Yahya Baykal, Optics \& Laser Technology, {\bf 42}, 1344 (2010).

\bibitem{ndagano2017optlet}
Bienvenu Ndagano, Nokwazi Mphuthi, Giovanni Milione, Andrew Forbes, Optics letters, {\bf 42}, 4175 (2017).

\bibitem{cox2018prapplied}
Mitchell A Cox, Ling Cheng, Carmelo Rosales-Guzm{\'a}n, Andrew Forbes, Physical Review Applied, {\bf 10}, 024020 (2018).

\bibitem{pang2018optlet}
Kai Pang, Haoqian Song, Zhe Zhao, Runzhou Zhang, Hao Song, Guodong Xie, Long Li, Cong Liu, Jing Du, Andreas F. Molisch, Moshe Tur, and Alan E. Willner, Optics letters, {\bf 43}, 3889 (2018). 

\bibitem{zhu2016optexp}
Long Zhu, Jun Liu, Qi Mo, Cheng Du, Jian Wang, Optics express, {\bf 24}, 16934 (2016).

\bibitem{krenn2014njp}
Mario Krenn, Robert Fickler, Matthias Fink, Johannes Handsteiner, Mehul Malik, Thomas Scheidl, Rupert Ursin, and  Anton Zeilinger, New Journal of Physics, {\bf 16}, 113028 (2014).

\bibitem{krenn2016pnas}
Mario Krenn, Johannes Handsteiner, Matthias Fink, Robert Fickler, Rupert Ursin, Mehul Malik, and  Anton Zeilinger, Proceedings of the National Academy of Sciences, {\bf 113},13648 (2016).

\bibitem{erhard2018LSA}
Manuel Erhard, Robert Fickler, Mario Krenn, and Anton Zeilinger, Light: Science \& Applications, {\bf 7}, 17146 (2018).


\bibitem{roychowdhury2004oc}
Hema Roychowdhury, and  Emil Wolf, Optics communications, {\bf 241} 11 (2004).

\bibitem{avramov2016oc}
S. Avramov-Zamurovic,  C. Nelson, S. Guth, Olga Korotkova, and  R. Malek-Madani, Optics Communications, {\bf 359}, 207 (2016).


\bibitem{ricklin2002josaa}
Jennifer C Ricklin, and Frederic M Davidson, JOSA A, {\bf 19}, 1794 (2002).
 
\bibitem{salem2003oc}
M. Salem, T. Shirai, A. Dogariu, and E. Wolf, Optics Communications, {\bf 216}, 261 (2003).

\bibitem{wang2017ao}
Minghao Wang, Xiuhua Yuan, and Donglin Ma, Applied optics, {\bf 56},2851 (2017).


\bibitem{chen2016apl}
Yahong Chen, Sergey A Ponomarenko, Yangjian Cai, Applied Physics Letters,  {\bf 109}, 061107 (2016).

\bibitem{cai2014josaa}
Yangjian  Cai, Yahong Chen, and Fei Wang, JOSA A, {\bf 31}, 2083 (2014).

\bibitem{chen2014optexp}
Yahong Chen, Fei Wang, Chengliang Zhao, Yangjian Cai, Optics express, {\bf 22}, 5826 (2014).

\bibitem{liang2018optlet}
Chunhao Liang, Xinlei Zhu, Chenkun Mi, Xiaofeng Peng, Fei Wang, Yangjian Cai, Sergey A Ponomarenko, Optics letters, {\bf 43}, 3188 (2018).

\bibitem{liang2014optlet}
Chunhao Liang, Fei Wang, Xianlong Liu, Yangjian Cai, and Olga Korotkova, Optics letters, {\bf 39}, 769 (2014).

\bibitem{mei2013optexp}
Zhangrong Mei, Elena Shchepakina, and Olga Korotkova, Optics express, {\bf 21},17512 (2013).

\bibitem{liu2016optlet}
Xianlong Liu,  Jiayi Yu,  Yangjian Cai, and Sergey A Ponomarenko, Optics letters, {\bf 41}, 4182 (2016).

\bibitem{zhu2017optexp}
Jie Zhu, Xiaoli Li, Huiqin  Tang, and  Kaicheng Zhu, Optics express, {\bf 25},20071 (2017).

\bibitem{yu2018optexp}
Jiayi Yu, Fei Wang, Lin Liu, Yangjian Cai, and Greg Gbur, Optics express, {\bf 26}, 16333 (2018).

\bibitem{lin2020optexp}
Rong Lin, Hancheng Yu, Xinlei Zhu, Lin Liu, Greg Gbur, Yangjian Cai, and Jiayi Yu, Optics Express, {\bf 28}, 7152 (2020).  

\bibitem{aarav2017pra}
Shaurya Aarav, Abhinandan Bhattacharjee, Harshawardhan Wanare, and Anand K Jha, Physical Review A {\bf 96} 033815 (2017).

\bibitem{mandel1995coherence}
L. Mandel and E. Wolf, {\em Optical Coherence and Quantum Optics} (Cambridge
  university press, New York, 1995).

\bibitem{fried1966josaa}
D. Fried, J. Opt. Soc. Am. A {\bf 56}, 1372 (1966).

\bibitem{boyd2011optexp}
Robert W. Boyd, Brandon Rodenburg, Mohammad Mirhosseini, and Stephen M. Barnett, Optics express, {\bf 19}, 18310 (2011).

\bibitem{goodman2005introduction}
Joseph W Goodman, {\em Introduction to Fourier optics}, ({Roberts and Company Publishers, Englewood, Colorado} 2005).

\bibitem{mandel1962measures}
Leonard Mandel, and Emil Wolf, Proceedings of the Physical Society, {\bf 80}, 894 (1962).

\bibitem{rodenburg2012optlet}
Brandon Rodenburg, Martin PJ Lavery, Mehul Malik, Malcolm N O’Sullivan, Mohammad Mirhosseini, David J Robertson, Miles Padgett, Robert W Boyd, Optics letters, {\bf 37}, 3735 (2012).

\bibitem{ponomarenko2007ol}
Sergey A Ponomarenko, Weihong Huang, and Michael Cada, Optics letters, {\bf 32}, 2508 (2007).

\bibitem{zhu2019ol}
Xinlei Zhu, Fei Wang, Chengliang Zhao, Yangjian Cai, and Sergey A Ponomarenko, Optics letters, {\bf 44}, 2260 (2019).

\bibitem{li2020ol}
Xiaofei Li, Sergey A Ponomarenko, Zhiheng Xu, Fei Wang, Yangjian Cai,  and  Chunhao Liang, Optics Letters, {\bf 45}, 698 (2020).

\bibitem{naraghi2017ol}
Roxana Rezvani Naraghi, Heath Gemar, Mahed Batarseh, Andre Beckus, George Atia, Sergey Sukhov,  and  Aristide Dogariu, Optics letters, {\bf 42}, 4929 (2017).

\end{thebibliography}

\end{document}